\documentclass[a4paper,superscriptaddress,twocolumn,prl]{revtex4}
\usepackage{graphicx}
\usepackage{amsmath}
\usepackage{hyperref}
\usepackage{xcolor}
\usepackage[]{natbib}
\usepackage{siunitx}

\begin{document}
\title{A quantum sensor: simultaneous precision gravimetry and magnetic gradiometry with a Bose-Einstein condensate}

\author{K.S. Hardman}
\email{kyle.hardman@anu.edu.au}
\homepage{http://atomlaser.anu.edu.au/}
\author{P.J. Everitt}
\author{G.D. McDonald}
\author{P.Manju}
\author{P.B. Wigley}
\author{M.A. Sooriyabandara}
\author{C.C.N. Kuhn}
\author{J.E. Debs}
\author{J.D. Close}
\author{N.P. Robins}

\affiliation{Quantum Sensors and Atomlaser Lab, Department of Quantum Science, Australian National University, Canberra, 0200, Australia}

\date{\today} 

\begin{abstract} A Bose-Einstein condensate is used as an atomic source for a high precision sensor.  A $5\times 10^6$\,atom F=1 spinor condensate of $^{87}$Rb is released into free fall for up to $750$\,ms and probed with a $T=130$\,ms Mach-Zehnder atom interferometer based on Bragg transitions.  The Bragg interferometer simultaneously addresses the three magnetic states, $\left| m_f=1,0,-1 \right\rangle$, facilitating a simultaneous measurement of the acceleration due to gravity with a 1000 run precision of $\Delta$g/g$=1.45\times 10^{-9}$ and the magnetic field gradient to a precision $120$\,pT/m. 
\end{abstract}

\maketitle

Acquiring accurate and precise data on magnetic and gravity fields is critical to progress in mineral discovery \cite{Lyatsky2010,Campbell1996}, navigation \cite{DeGregoria} and climate science \cite{Chen2006}.  A diverse array of tools and techniques have been developed to improve the quality of measurements, including macroscopic classical springs \cite{Lacoste1967} falling corner cubes \cite{Faller1986}, solid-state \cite{Merayo200571} and superconducting \cite{Goodkind1999} systems. Following the early pioneering work in precision atom interferometry \cite{Peters2001}, the past decade has seen devices using cold atomic sources become competitive with traditional acceleration sensors \cite{kasevichOld, MullerGravityTest, LeGouët2008, Geiger_airbourne, wuhan2013}.  Technical developments improving size, weight and power have allowed for applications in space science \cite{Altschul2015501} and field ready state-of-the-art gravimeters and gradiometers \cite{muquans, atomsensors, aosense}.  

Like their classical counterparts, sensors based on cold atoms measure the trajectory of the test particles \cite{Schmidt}. Unlike classical particles, atoms offer internal degrees of freedom, allowing for the possibility of additional simultaneous measurements including time, magnetic fields and magnetic field gradients.  Although these advantages are intrinsic to all atomic sources, ultra-cold Bose-Einstein condensates (BEC) offer additional benefits over thermal atoms.  An intrinsic feature of a BEC is a spatial coherence equivalent to the size of the cloud which is generally $100$'s of $\mu$m while thermal sources have spatial coherence length on the order of the de Broglie wavelength, $\lambda_{dB}=\sqrt\frac{2 \pi \hbar^2}{m k_B T}$ ($\sim$ 0.1$\mu$m).  This spatial coherence has been shown to provide robustness to systematics which result in loss of fringe contrast such as cloud mismatch at the final beam splitter pulse \cite{PhysRevA.89.023626}.  The BEC then allows a sensor to be operated unshielded in varying environments where background field gradients and curvatures are non-negligible.

This letter introduces a new type of sensor which {\em simultaneously} measures gravity and magnetic field gradients to high precision.  In this lab based sensor, an optically trapped cloud of $^{87}$Rb atoms is cooled to condensation and projected into an $F=1$ spin superposition, then passed through a vertical light pulse Mach-Zehnder interferometer based on Bragg transitions \cite{DebsBECgrav,OurGravimeter}.  The spin superposition in combination with the large spatial coherence of the BEC allows simultaneous precision measurement of gravity and absolute magnetic field gradient in an unsheilded device \cite{Bloch:2000yj}. A $2\times 10^6$ atom condensate is used in the interferometer \cite{1367-2630-14-2-023009} with no loss in contrast over all interferometer times.  

\begin{figure}
\centering\includegraphics[width=240pt]{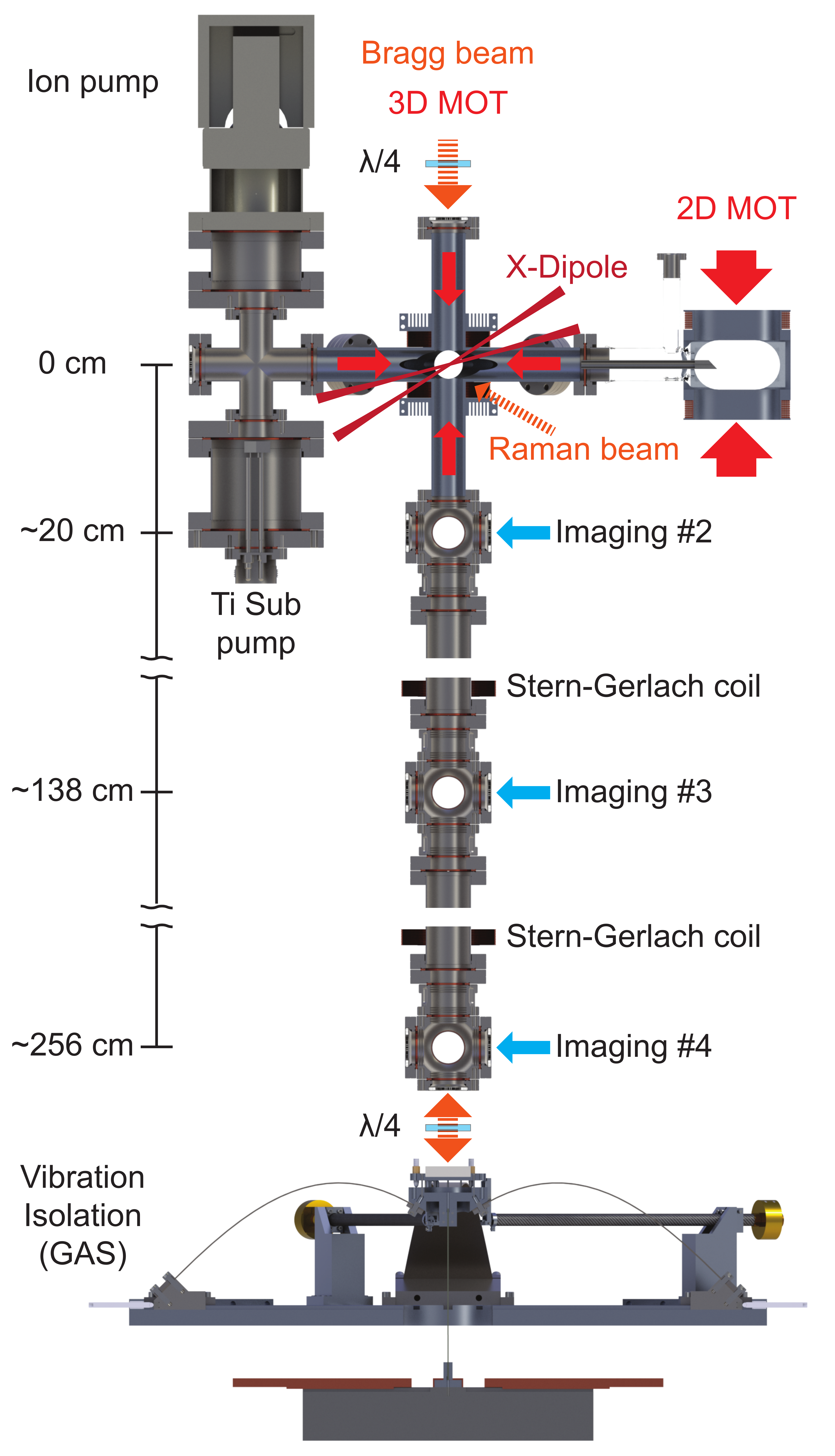}
\caption{Overview of the experimental setup.}
\label{figapparatus}
\end{figure}

The experimental schematic is shown in Figure \ref{figapparatus}.  A hot sample of $^{87}$Rb atoms is created and precooled in a $2$D magneto-optical trap (MOT). These precooled atoms are transferred to an aluminum ultra-high vacuum cell via a high impedance gas flow line and blue detuned push beam.  In $6$\,s, $5\times 10^9$\,atoms are collected in a $3$D MOT where a standard compression and polarization gradient cooling sequence is applied achieving a $20$\,$\mu$K temperature.  The atoms are then loaded into a hybrid magnetic quadropole and crossed optical dipole trap.  An initial stage of evaporation is completed using a microwave knife over $4.5$\,s leaving $4\times 10^7$\,atoms at $4$\,$\mu$K and a phase space density of $1\times 10^{-4}$.  The magnetic field gradient is decreased from $150$\,G/cm to $25$\,G/cm over $200$\,ms where the atoms are no longer supported against gravity. This efficiently loads all the atoms into a crossed dipole trap.  The magnetic field is subsequently reduced to zero and extinguished.  A pair of $1070$\,nm broad linewidth fiber lasers intersecting at $22.5^\circ$ each with beam waists of $300$\,$\mu$m provide the optical trap volume.  Forced evaporation is then completed by simultaneously reducing the intensity of both optical beams.  After $2$\,s of evaporation a pure  $\left| F=1, m_f=-1\right\rangle $ $2\times 10^6$\,atom condensate is formed with an in trap width of $\sim 50$\,$\mu$m.  The effective thermal temperature of the condensate is estimated to be significantly below $50$\,nK for up to $750$\,ms expansion.  Controlling the final trap depth allows for the creation of a pure BEC or ultra-cold thermal source.  Following production of the condensate, the optical trap is extinguished suddenly ($\sim$10\,$\mu$s) and the atoms fall under gravity.  The apparatus allows for a $750$\,ms time-of-flight (TOF) with four regions available for horizontal imaging at $0-25$\,ms, $220$\,ms, $530$\,ms, and $750$\,ms.  Standard absorption imaging techniques are used for the $0-25$\,ms and $220$\,ms imaging regions.  Frequency modulation imaging (FMI) \cite{Bjorklund1983, HardmanFMI} is implemented for the $530$\,ms and $750$\,ms drop times used for high sensitivity interferometers.

\begin{figure}
\centering\includegraphics[width=250pt]{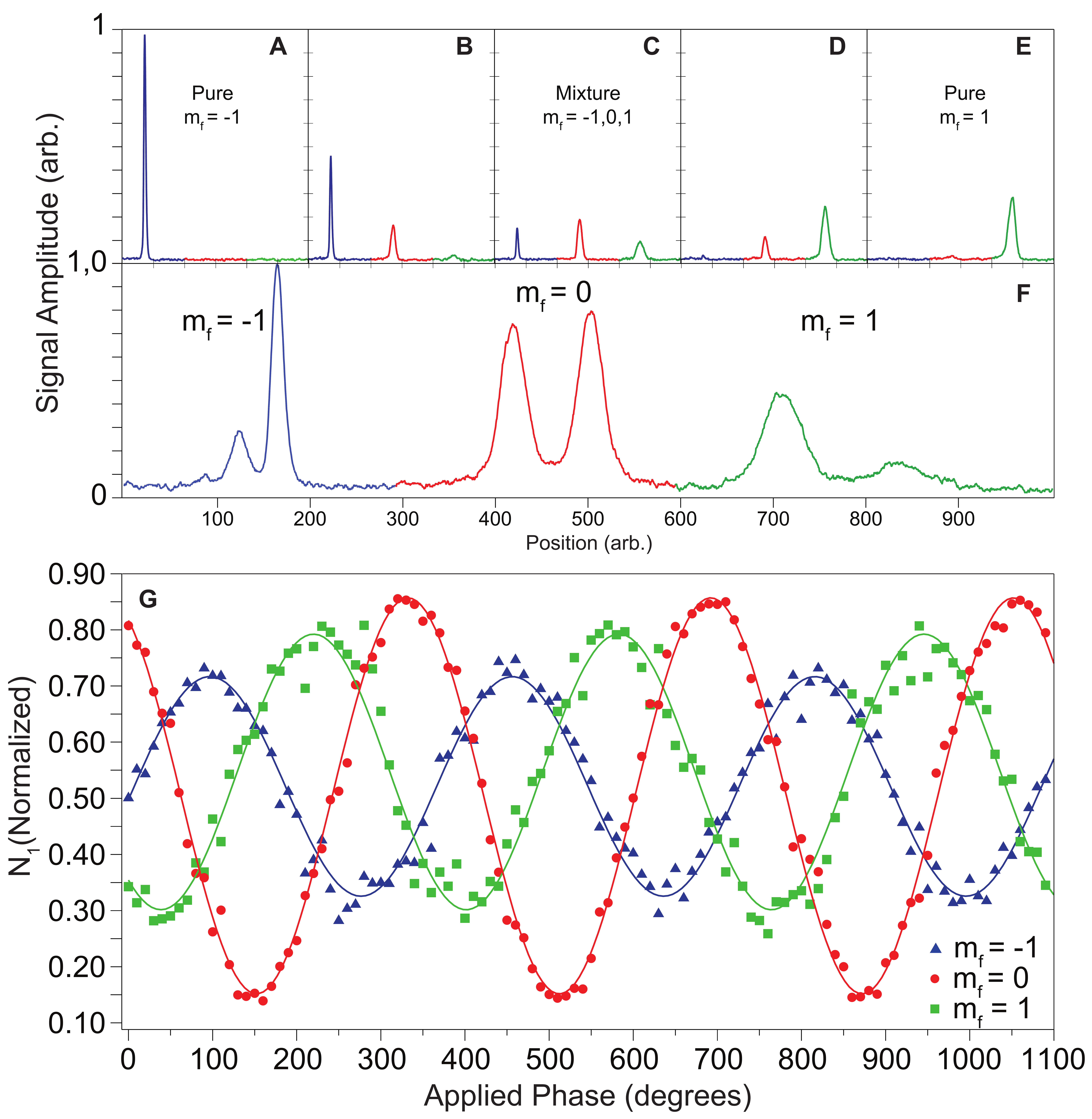}
\caption{A-E) By adjusting Raman pumping power it is possible to produce pure $\left| m_f=-1 \right\rangle$ (A), $\left| m_f=1 \right\rangle$ (E) or any spin mixture condensate.  A Stern-Gerlach pulse separates the states before imaging.  The variation in cloud profile is due to magnetic lensing from curvature in the Stern-Gerlach field.  F) A typical FMI image of a three state interferometer.  G) The interference fringes produced by a $T=60$\,ms Mach-Zehnder interferometer with a spinor BEC source achieving $9$\,mrad phase noise in $100$\,runs. Changes in local gravity are monitored through correlated phase shifts of all $m_{\text{f}}$ states.  The absolute background field, B, and field gradients, dB/dz, are found from the relative phase shifts between all states ($\Delta \Phi_{-1}$, $\Delta \Phi_{1}$, and $\Delta \Phi_{-1}+\Delta \Phi_{1}$)using equation \ref{eqnBR}.  Changes in dB/dz are monitored through anti-correlated phase shifts of $\left| m_{\text{f}}=-1\right\rangle$ and $\left| m_{\text{f}}=1\right\rangle$ relative to $\left| m_{\text{f}}=0\right\rangle$.}
\label{figintFMI}
\end{figure}

A pair of far detuned, co-propagating, mutually linear polarized beams are pulsed on 5\,ms after the atoms are released from the optical trap to induce Raman transitions.  These transfer the condensate into a magnetic superposition state of $\left| m_f=1,0,-1 \right\rangle$, Figure \ref{figintFMI}(A-E).  The Raman beams are aligned through mutual fiber coupling ensuring negligible relative angle.  A vertically oriented Mach-Zehnder atom interferometer consisting of three $30$\,GHz detuned Bragg pulses (ensuring orthogonal internal $m_f$ states) is used to measure the phase accumulated on all three internal states as they fall.  The Bragg lattice is generated by two frequency shifted beams with orthogonal linear polarizations.  These beams are coupled to the apparatus head in a single polarization-maintaining single-mode fiber and then passed through a quarter waveplate ($\lambda/4$) before and after the atoms.  This is followed by the inertial reference retro reflector.  The orientation of Bragg optics is such that the Bragg transitions are driven by circularly polarized light.  One beam is frequency chirped to match the increasing doppler shift of the atoms as they fall, while the other is adjusted to address the resonance frequency for transfer of $2$\,$\hbar$k of momentum, where $k$ is the wavenumber of the light.  The Bragg laser is aligned to vertical using a liquid mercury mirror and back coupled into the output fiber over a $6$\,m total path length.  A home built external cavity diode laser seeds a frequency doubled $1560$\,nm fiber laser system capable of producing $11$\,W of $780$\,nm light with a linewidth of $5$\,kHz, generating the Bragg and Raman light \cite{Sane:12}.  Independent frequency control and pulse shaping for all Bragg and Raman beams is accomplished using four acousto-optic modulators (AOMs) driven with a direct digital synthesizer (DDS) and referenced to a Cesium primary frequency standard.  No active optical phase locking system has been implemented in the current setup. 

A $5$\,cm gold mirror mounted on a home built geometric anti-spring (GAS) \cite{Cella2005} provides the inertial reference.  The GAS provides passive filtering of ground vibrations by virtue of a low frequency mechanical oscillator.  This passive oscillator is tuned to an ultra-low natural frequency $180$\,mHz.  A direct measurement of the GAS transfer function shows significant isolation from $1$\,Hz ($<-22$\,dB) and greater ($<-76$\,dB at $70$\,Hz).

\begin{figure}
\centering\includegraphics[width=250pt]{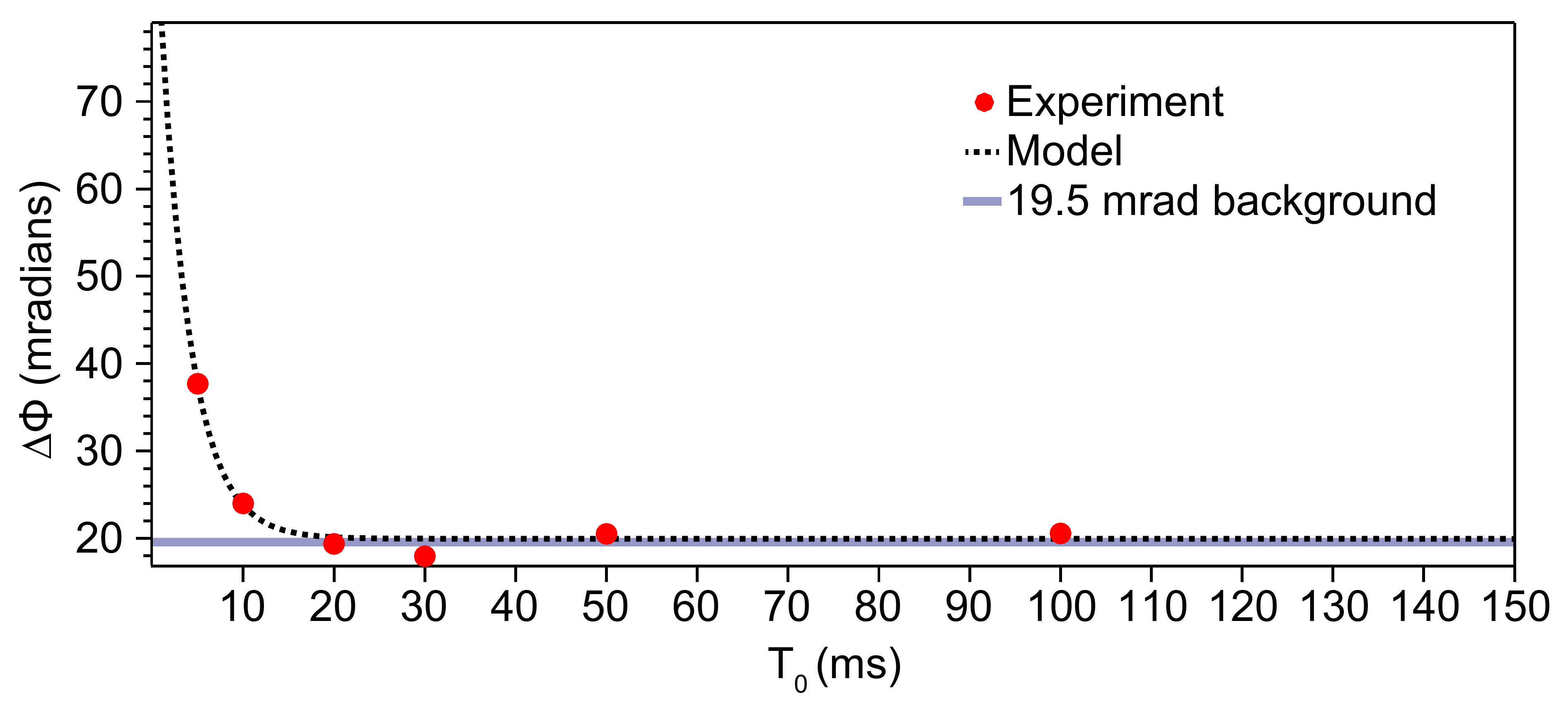}
\caption{A pure BEC source is allowed to expand for various times, $T_0$, before a $T=60$\,ms Mach-Zehnder interferometer is performed.  The efficiency of the first beam splitter pulse is varied by up to $15$\,$\%$ ensuring different atom numbers in each arm.  This number imbalance will lead to a relative phase shift on the arms due to mean field interactions.  By performing many runs the phase noise from this number imbalance and initial cloud density can be found.  The phase noise from mean field effects in this system asymptote at $20$\,ms to the background phase noise from the Bragg laser system.  A basic mean field model of the system is plotted against this data with good agreement.}
\label{figmeanfield}
\end{figure}

The various experimental cycles and data acquisition are as follows.  A BEC of $~2\times 10^6$\,atoms is created and released into free fall.  $5$\,ms after release the atoms are either left in the initial $m_f$ state or transferred into a superposition of $~25$\,$\%$ $\left| m_f=-1\right\rangle $, $50$\,$\%$ $\left| m_f=0\right\rangle $, and $25$\,$\%$ $\left| m_f=1\right\rangle $, Figure \ref{figintFMI}(C).  The time between cloud release and the start of the interferometer, $T_0$, is varied from $7$ to $100$\,ms to investigate the effect of undissipated mean field energy on the phase noise of the interferometer, Figure \ref{figmeanfield}. 
 This phase noise is seen to be correlated to cloud density variations due to imperfect beam splitter pulses and asymptotes at $T_0\approx 20$\,ms.  This behavior follows a simple mean field model of energy \cite{DebsBECgrav} and corresponds to $\sim$1\,mrad additional phase noise at $T_0\approx 20$\,ms.  Due to this effect $30$\,ms of free expansion is allowed for all high sensitivity interferometers.  
 
\begin{figure}
 \centering\includegraphics[width=250pt]{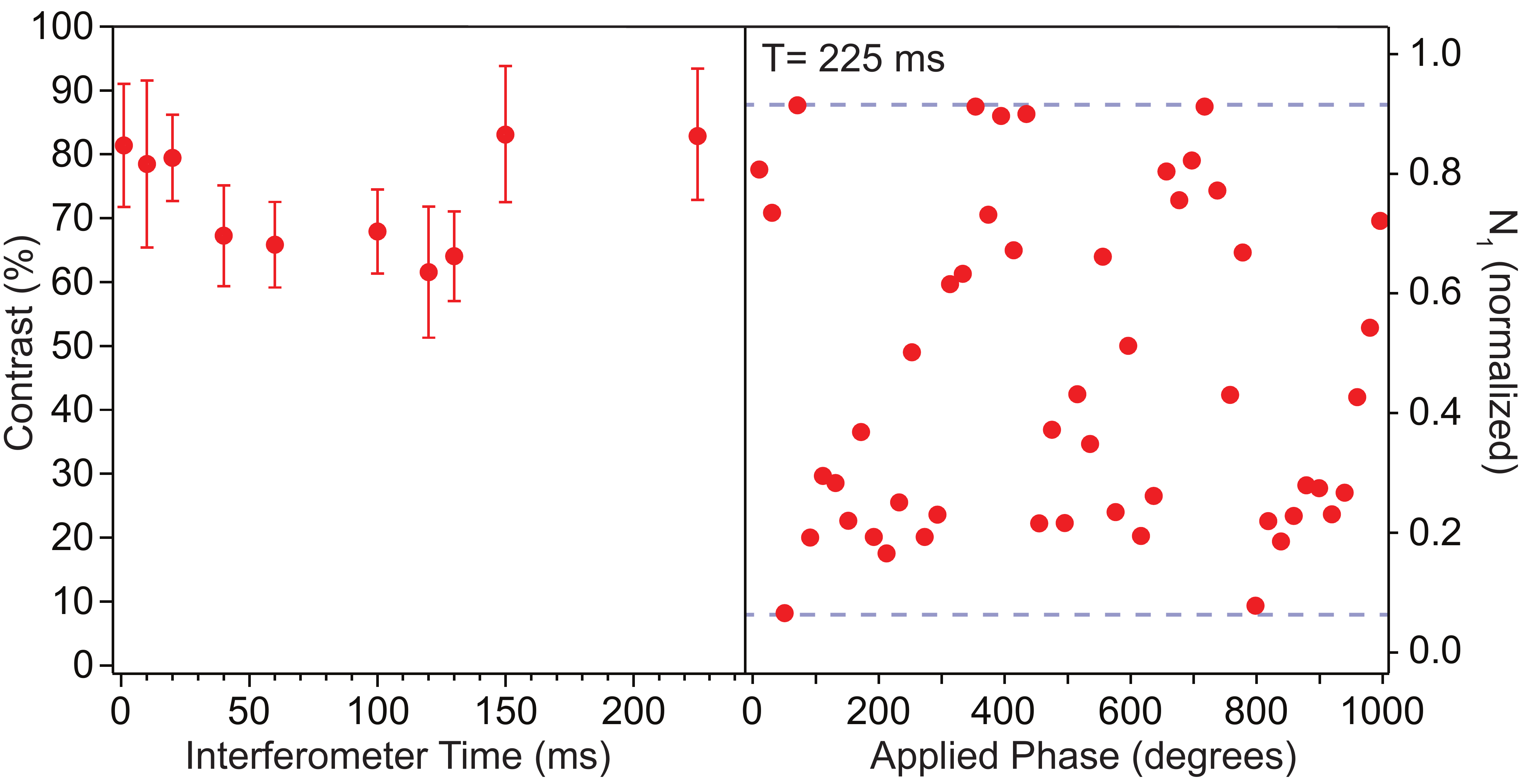}
 \caption{The contrast of a $\left| m_{\text{f}}=-1\right\rangle $ BEC interferometer for varying $T$.  Contrast is defined here as the difference between the maximum and minimum of an interference fringe.}
  \label{figcontrast}
 \end{figure}

 Following free expansion a sequence of Gaussian shaped Bragg pulses of $50$\,$\mu$s full width half maximum are applied to form a Mach-Zehnder interferometer.  By scanning the phase of the final beam splitter an interference fringe can be produced (Figure \ref{figintFMI}(G)).  The time between interferometer pulses, $T$, may be varied from 1 to 250\,ms.  By scanning $T$ in a magnetically sensitive interferometer the fringe contrast is explored in regions of varying cloud separation at the final beam splitter, Figure \ref{figcontrast}. 
  The curvature of the magnetic field along the $2.5$\,m drop leads to unequal accelerations felt by each interferometer arm.  This asymmetric acceleration results in a final positional separation of the two arms at the final beam splitter pulse.  The total separation is proportional to $T^2$.  A BEC source is shown to maintain contrast ($\approx 80$\,$\%$) over the full interferometer range suggesting that the coherence length of a BEC ensures robustness to systematics leading to mismatch at the final beam splitter.  In comparison a thermal cloud (velocity selected to $90$\,nK) undergoing the same experimental sequence exhibits $7$\,$\%$ contrast at $T=40$\,ms.   Following the final beam splitter pulse a variable amplitude magnetic field gradient from a vertical co-axial solenoid can be applied to impart Stern-Gerlach separation of the different magnetic states (Figure \ref{figintFMI}). The $2$\,$\hbar$k momentum separation of the interferometer arms requires $200$\,ms of free propagation for sufficient separation of the final momentum states. Large momentum transfer beamsplitters would alleviate the need for this long separation time \cite{PhysRevA.88.053620}. Finally, the number of atoms in each momentum and internal state is measured using FMI.  The total phase shift accumulated in each interferometer is given by $\phi_{\text{total}}=\phi_{\text{g}}+\phi_{\text{B}}$ where $\phi_{\text{g}}$ and $\phi_{\text{B}}$ are given by: 
   \small
       \begin{equation}
       \label{eqnBR}
       \begin{aligned}
      &\phi_{\text{g}}=nk_{\text{eff}}gT^{2}\\
      &\phi_{\text{B}}=\frac{nk_{\text{eff}}T^{2}}{m_{\text{Rb}}}\frac{\partial B}{\partial z}\Bigg(g_{I}m_{f}-\frac{(g_{J}-g_{I})}{2\sqrt{A}}\bigg(\frac{m_f}{4}+x\bigg)\Bigg)
       \end{aligned}
       \end{equation}
       \normalsize

  $\phi_{\text{B}}$ is derived from the Breit-Rabi formula where $A=1+\frac{4m_{f}x+x^{2}}{4}$, with $x=\frac{(g_{J}-g_{I})}{\Delta E_{hfs}}\mu_{B}B$.  A typical three state FMI interferometer signal and scanned fringe is shown in Figure \ref{figintFMI}(G).  
  
\begin{figure}
 \centering\includegraphics[width=250pt]{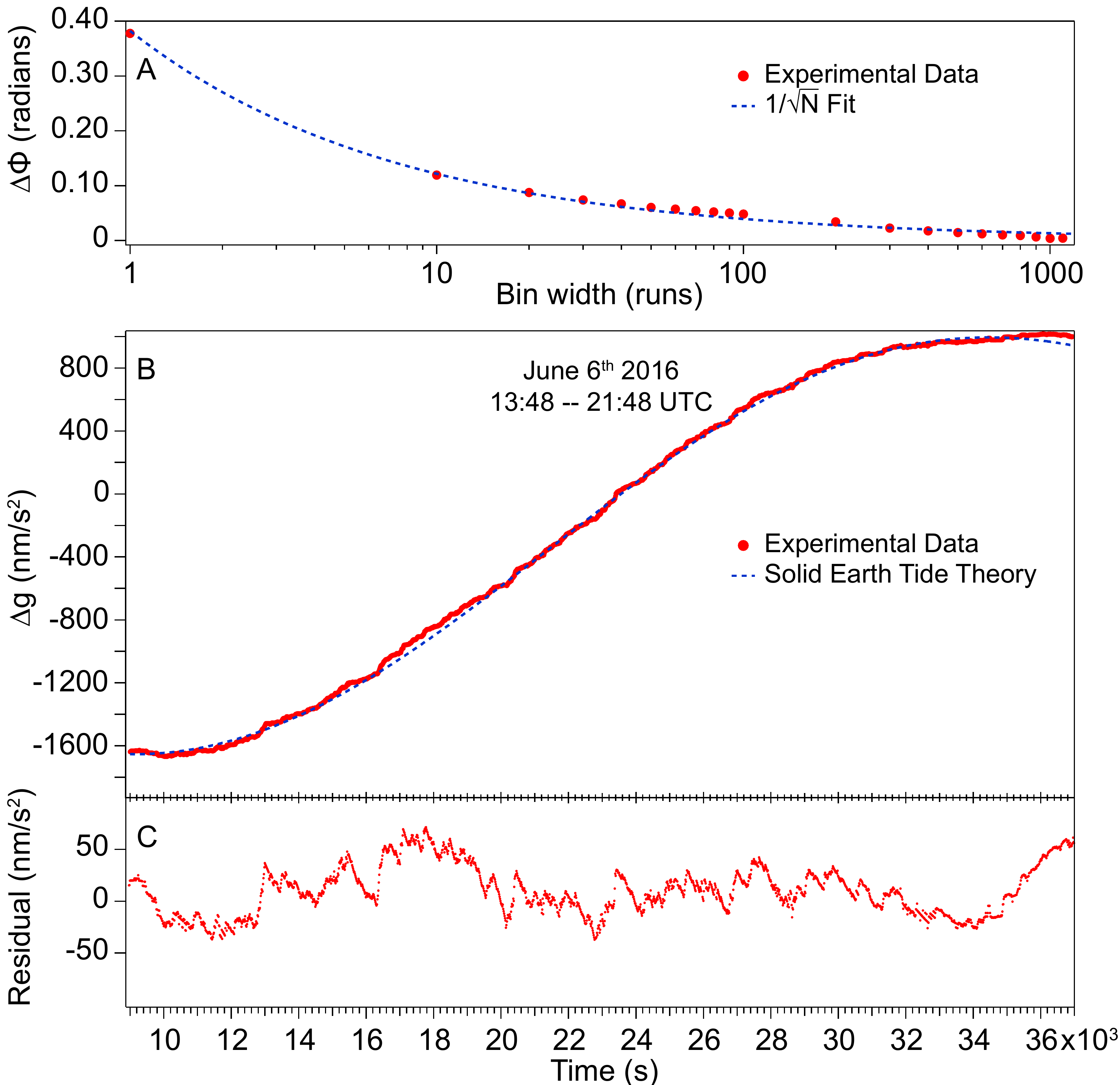}
 \caption{The deviation in gravitational acceleration over an 8 hour period is monitored using a spinor BEC sourced T=130\,ms interferometer. Data corresponding to the $\left| m_{\text{f}}=0\right\rangle $ state is plotted.  A) The integrated phase sensitivity of the interferometer corresponding to varying bin widths (circles) shown with a $1/\sqrt{N}$ fit (dashed).  A maximum phase sensitivity of $3.8$\,mrad is reached for a 1000\,run bin width.  B) 1000\,point running average of experimental data (circles) with the solid earth tide theory overlaid (dashed).  C) The residual of the experimental data and the solid earth tide theory. A maximum precision of $\Delta$g/g$=8\times 10^{-8}$ per run and $\Delta$g/g$=1.45\times 10^{-9}$ in 1000\,runs was achieved.}
\label{fighighsense}
\end{figure}

  A $T=130$\,ms spinor BEC interferometer was run continuously over an 8 hour period on June $6^{th}$ 2016 monitoring deviations in gravitational acceleration from the solid earth tides.  The integrated phase sensitivity of the apparatus follows the expected $1/\sqrt{N}$ scaling where N is the number of binned points (Figure \ref{fighighsense}(A)).  At $N=1000$ the sensitivity asymptotes to $3.8$\,mrad.  Figure \ref{fighighsense}(B) shows the experimental data with a 1000 point running average overlaid with the theoretical solid earth tide.  The residual of the experimental data and solid earth tide theory is shown in Figure \ref{fighighsense}(C).  The achieved gravitational acceleration sensitivity of this device reached a combined three state precision of $\Delta$g/g$=8\times 10^{-8}$ per run and $\Delta$g/g$=1.45\times 10^{-9}$ in 1000\,runs.  This sensitivity is limited by laser phase noise on the passive Bragg laser system.  

As seen in equation \ref{eqnBR} and Figure \ref{figintFMI} the interferometers in the $\left| m_{\text{f}}=-1\right\rangle $ and $\left| m_{\text{f}}=1\right\rangle $ states are phase shifted with opposite signs away from the magnetically insensitive $\left| m_{\text{f}}=0\right\rangle $ state.  By comparing the relative phase shifts the three fringes can be combined to extract absolute magnetic field gradient, absolute magnetic field, and local gravity.  In addition, it is straightforward to extend the system to measure gravity gradients \cite{OurGravimeter}.
\begin{figure}
  \centering\includegraphics[width=250pt]{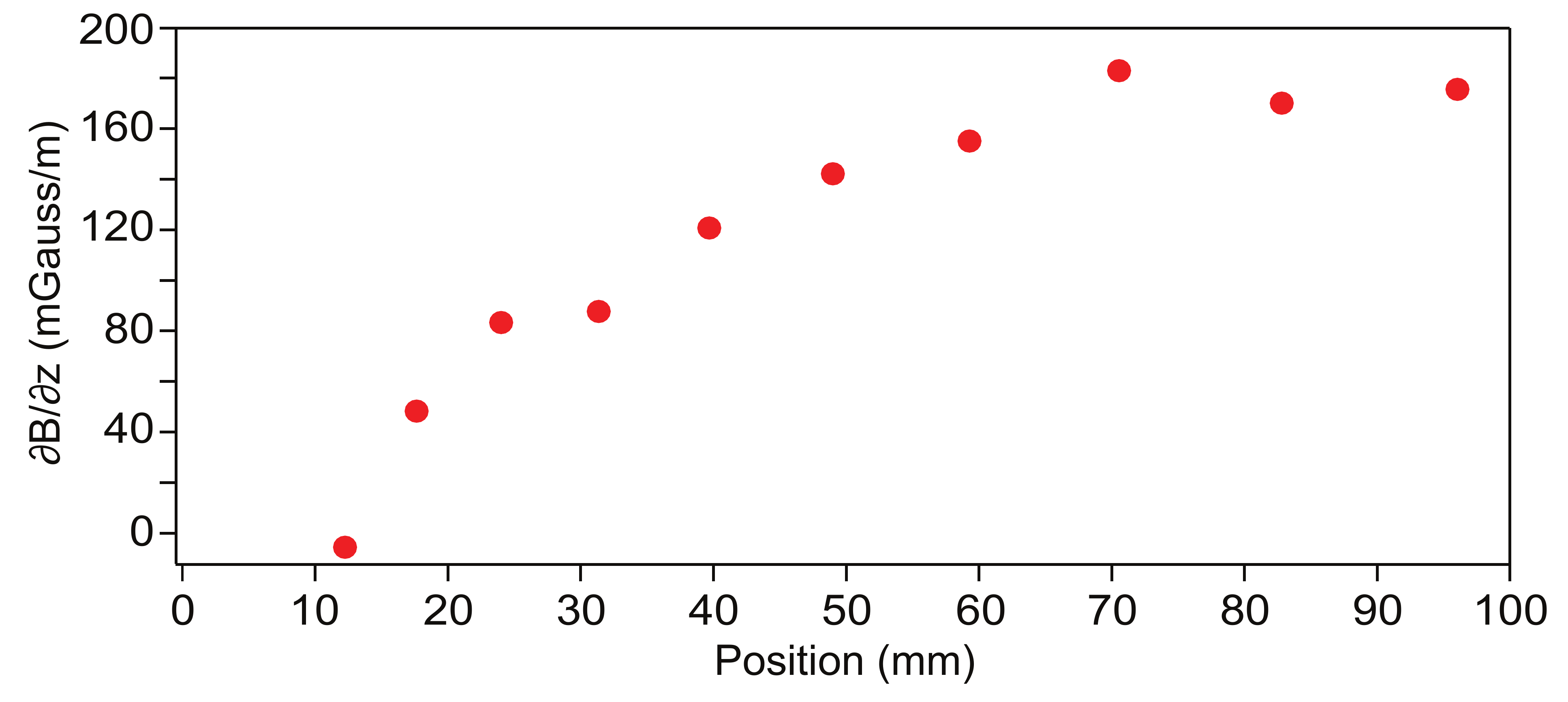}
  \caption{The magnetic field gradient along a portion of the drop in mapped by varying the initial position of a $T=40$\,ms interferometer where the initial cloud has been prepared in a superposition of $\left| m_f=-1\right\rangle $ and $\left| m_f=1\right\rangle$.  The error in gradient measurement is less than the point size.}
  \label{figdbdz}
  \end{figure}
  
  Figure \ref{figdbdz} shows the measure of magnetic field gradient along a portion of the drop. 
 The maximum achieved phase sensitivity enables an absolute magnetic field gradient sensitivity of $120$\,pT/m. To date this is the highest achieved atom based magnetic gradient sensitivity.  This is competitive with the state of the art relative devices such as solid state ($0.52$\,nT/m, 1 second integration) \cite{Jun2014} and SQUID based systems ($10$\,pT/m, 1 second integration) \cite{Du2010}.  A more sophisticated analysis of the sensor noise is likely to achieve significantly higher precision through common mode noise reduction.    At current precision this sensor is capable of recognizing weak magnetic features such as paramagnetic rock anomalies at depths of $1$\,km  \cite{Merayo200571}.   This step change in sensor technology enabled by fundamental properties in source selection will allow for the simultaneous precision exploration of gravitational and magnetic anomalies leading to higher spatial resolution mapping as well as the ability to differentiate the feature's material properties.

In conclusion, a high precision simultaneous gravimeter, magnetic gradiometer, and magnetometer based on a free falling Bose-condensed source has been demonstrated.  The atomic source provides internal degrees of freedom allowing the preparation of magnetically sensitive states.  Due to the large and spatially varying background magnetic field of the surrounding environment the simultaneous measurement of all states (magnetic and non-magnetic) would not be possible without the macroscopic spatial coherence provided by the condensed source.  The flexibility of this device allowed for direct mean-field noise characterization of a BEC in a high precision apparatus.  Furthermore a direct comparison of thermal and BEC sources was achieved showing that under near identical systematic conditions a condensate has contrast a factor of 5 higher than the thermal cloud.  A full noise characterization of the system will be investigated in the future.  This is the first iteration towards an all-in-one quantum sensor which will be capable of simultaneous measurement of g, $\Delta$g, B, $\Delta$B, rotations, and time. Currently work is ongoing to implement symmetric horizontal Bragg transitions to also measure rotations. 

A technical hurdle that needs to be overcome for condensed sources to become viable in field deployable devices is the time required to prepare a large BEC ($10$\,s).  Significant progress towards solving this problem has been reported and summarized recently \cite{1367-2630-17-6-065001}.  Nonetheless, fast condensate production generally sacrifices total atom number for duty cycle.  This is evident from the best integrated flux \cite{1367-2630-17-6-065001} achieved in fast devices, $2.5\times 10^5$\,atoms/s, when compared to the $4\times 10^5$\,atoms/s flux in this device.  The application of techniques such as sideband cooling offer a path to improving flux on both atom chip and free-space based sensors \cite{PhysRevLett.108.103001} without sacrificing atom number.  The possibility for the integration of sideband cooling techniques with the current sensor are being explored.

\bibliographystyle{apsrev_v2}
\bibliography{ref}
\end{document}